\documentclass{elsarticle}
\usepackage[T1]{fontenc}
\usepackage{amsfonts}
\usepackage{amssymb}
\usepackage[latin9]{inputenc}
\usepackage{graphicx}

\begin{document}

\begin{frontmatter}

\title{An Opinion Dynamics Model for the Diffusion of Innovations}
\author{Andr\'e C. R. Martins, Carlos de B. Pereira and Renato Vicente}
\address{GRIFE-EACH, Universidade de S\~ao Paulo, Campus Leste, 03828-080, S\~ao Paulo-SP, Brazil}

\begin{abstract}
We study the dynamics of the adoption of new products  by agents with continuous opinions and discrete actions
(CODA). The model is such that the refusal in adopting a new idea
or product is increasingly weighted by neighbor agents as evidence
against the product. Under these rules, we study the distribution of adoption times 
and the final proportion of adopters in
the  population. We compare the cases where initial adopters
are clustered to the case where they are randomly scattered around the social network and 
investigate small world effects on the final proportion of adopters.
The model predicts a fat tailed distribution for late adopters which is verified by empirical data.
\end{abstract}
\end{frontmatter}

\section{Introduction}

Social agents can be mutually influenced by exchanging information
on their opinions about a set of issues of interest. This scenario
has been studied by a number of Opinion Dynamics models~\cite{bordognaalbano07,castellanoetal07,galametal82,galammoscovici91,
sznajd00,stauffer03a,deffuantetal00,hegselmannkrause02}.
Some of them suppose opinions to be discrete~\cite{galammoscovici91,sznajd00,stauffer03a},
while others treat opinions as continuous~\cite{deffuantetal00,hegselmannkrause02}. 
A third type introduces a dichotomy between internal opinions and discrete actions with dynamical
consequences that must not be understated~\cite{urbig03,martins08a,martins08b,vicenteetal08b}.
One should note that opinions studied in those problems can represent
political issues or the quality evaluation of some new idea or product~\cite{galamvignes05}.

The marketing literature regards products (goods, services or ideas)
as new if they are unknown within a particular market ~\cite{kotlerarmstrong2003}.
An \textit{innovation} is defined as a product that is new in each
and every market. When an innovation is adopted by consumers, the
launcher firm may obtain a competitive advantage for a considerable
period of time as  legal protection against copies (patents) are
usually in place.

The identification of factors leading to a successful innovation in
a given market is a question of considerable practical importance.
It seems clear that this question may be approached either by delving
into issues concerning product launching or by considering the dynamics
of innovations adoption~\cite{wejnert02}. To date the literature on marketing has mainly
focused on innovation analysis considering variables under the strict
control of the launcher firm. However, we still lack understating
of the dynamics followed by innovations adoption, despite the clear
relevance of the subject.

In order to describe the process of innovation adoption, practitioners \cite{kotlerarmstrong2003,solomon2002,tiddetal2005}
employ a heuristic model, proposed in 1962 by Rogers~\cite{rogers2003},
that summarizes  evidence gathered by a number of authors
on diverse markets. This model describes the adoption process as following
a well defined dynamics with consumers differing in their predisposition
to buy. At first, a small number of consumers buys the innovation,
sales increase as influence spreads until a saturation level and, finally,
the number of new adopters declines as the last consumers with interest
on the product are reached~\cite{bass69a,rogers2003}. The resulting  distribution of adoption times 
is usually described as being a  ``normal curve''.   Based on these Rogers'
innovation adoption curves, practitioners divide by convention
consumers into five groups (segments): innovators, early adopters, early majority,
late majority and laggards (late adopters). Each segment is supposed
to be composed by homogeneous life-styles which are described following
survey studies. Observational studies on innovation diffusion
usually postulate these five segments, thus trying to
fit consumers into rigidly defined  classes~\cite{beaudoinetal03,valente96a}.
A number of authors, therefore, seek the \emph{ad hoc} identification
of a group of innovative consumers that should correspond to the $2.5\%$
earliest adopters in the targeted market~\cite{goldsmithflynn92}.
The same sort of heterogeneity is also supposed to hold when agent-based
models are considered ~\cite{garcia05a}.

We propose a simulation of the innovation adoption process without
assuming an a priori classification of consumers into segments. In
the model we put forward agents  assign  probabilities to the idea
that they should adopt an innovation and then decide accordingly. Probabilities
are then updated by observing decisions of neighboring agents. Since
the opinion is a continuous variable (a probability) and observations
are discrete (adoption of a product), the model is a slightly modified
version of the Continuous Opinions and Discrete Actions (CODA) model
~\cite{martins08a,martins08b}. The CODA  model is modified by introducing dynamic likelihoods that  allow the importance of distinct observations to change in time.

In this paper we have used simulations to find adoption curves as an emerging
property of the market without the need to rely on the usual {\it ad hoc}
segments. The resulting  curves are shown to be qualitatively similar
to Rogers' normal curves~\cite{rogers2003},  a power
law distribution for adopting times is observed for late adopters though. 

We  have  found that consumer expectancies on the time spent by the population to
 test an innovation  determines the final proportion
of adopters observed. Additionally, we have studied the dependency of the adopting proportion
on the initial distribution of early adopters and on the social network topology.
The diffusion process presented here evolves to a state where frozen interfaces separating adopters from non-adopters and  preventing further innovation diffusion emerge. We have also compared  curves for aggregate adoption as  predicted by the model to data from an observational study on the diffusion of a medical diagnostic innovation~\cite{poulsen98}.  

This paper is organized as follows. In the next section we propose a variation of the CODA model
 as a possible explanation to the diffusion of innovations process. In Section 3 
we describe the result of simulations on social networks defined by two dimensional regular lattices   with periodic boundary conditions and  edges rewired with probability $\lambda$. We analyze the cases of random and clustered initial locations for the early adopters. Section 4 presents a non-rigorous comparison of a  model  prediction with empirical data. Finally, a discussion of our results and  further directions are presented in a closing section.  

\section{The CODA model for the diffusion of innovations}

In the CODA model, agents update continuous opinions by observing  discrete choices taken by neighboring agents. We consider
two choices representing the adoption ($A$) or non-adoption ($B$)
of an innovation. To these binary choices we associate Ising spins $s_i=+1$ (A) or $s_i=-1$ (B). Each agent $i$ is provided with a subjective probabilistic opinion $p_i$ assigned to the proposition ``$A$ \textit{is the best choice that can be made}'' (and, therefore, a probability of $1-p_i$ assigned to the proposition  ``$B$ \textit{is the best choice that can be made}''). We emphasize that probabilities are employed here in the  Bayesian sense, namely, as a subjective belief on the truth of a particular assertion~\cite{jaynes03}.

 We suppose that agents always act according to the assertion  they believe most likely to be true. The discrete choice $s_i$ is, therefore, a step function of $p_i$. Namely, if $p_i>0.5$, agent $i$ chooses to be an adopter and its state is set to $s_{i}=+1$ (analogously, a non-adopter is set to state $s_{i}=-1$). 

Subjective beliefs (quantified by probabilities) are learned upon social interactions with neighboring agents. Learning  comes from the belief that other agents behave rationally, to say, that agents always prefer the option they consider the best. In order to use Bayes theorem~\cite{jaynes03}, we also need a likelihood, that is, the probability associated with the observed choices, assuming that a given assertion ($A$ or $B$) is true. Consider $\alpha=P(OA|A)$ to be the probability
of observing the adoption of the product by a neighbor ($OA$),
given that $A$ is true ( ``$A$ \textit{is the best choice that can be made}''). Similarly, consider
$\beta=P(OB|B)$ to be the probability of observing
a neighbor that is a non-adopter ($OB$), given that $B$ is true.

The belief of agent $i$ on $A$ at time $n$ is encoded into the (posterior) probability updated at every observation of neighboring agents $p_i(n)=P(A|\mathcal{O}_i(n))$, where $\mathcal{O}_i(n)$ stands for the particular sequence of observations made by agent $i$ up to interaction $n$. Bayes theorem prescribes how this update can be implemented ~\cite{ohagan94a}:  
\begin{equation}
P(A|\mathcal{O}_i(n))=\frac{ P(O_n|A) P(A|\mathcal{O}_i(n-1))}{P(O_n|A) P(A|\mathcal{O}_i(n-1)) + P(O_n|B)P(B|\mathcal{O}_i(n-1)) } ,\label{eq:bayes}
\end{equation} 
where $P(O_n|A)$ represents the likelihood of observation $O_n$ given that $A$ is true.

Along the lines of \cite{martins08a}, rational decisions can be made by looking at  log-odds defined as 
\begin{equation}
\nu_{i}(n)=\log\frac{P(A|\mathcal{O}_i(n))}{P(B|\mathcal{O}_i(n))}=\log\frac{p_i(n)}{1-p_i(n)}   .\label{eq:logodds}
\end{equation}
Log-odds $\nu_i$ are defined in the interval $-\infty<\nu_i<+\infty$ and can be regarded as a continuous local field  
over site $i$, the choice of agent $i$  being defined by a spin variable $s_i(\nu_i)=sign(\nu_i)$.

The Bayesian update   of agent $i$ belief upon the observation of its social neighbor's $j$ choice, as described by Equation \ref{eq:bayes}, yields the following simple prescription for  log-odds: 

\begin{equation}
\nu_{i}(n+1)=\left\{ \begin{array}{cc}
\nu_{i}(n)+a, & \textnormal{if }s_{j}=+1\\
\nu_{i}(n)-b, & \textnormal{if }s_{j}=-1,\end{array}\right.\label{eq:dynamics}
\end{equation}
 with
\begin{equation}
\begin{array}{ccc}
a & = & \log\frac{\alpha}{1-\beta}\\
b & = & \log\frac{\beta}{1-\alpha}.\end{array}\label{eq:logchange}
\end{equation}
Considering an asynchronous dynamics, at each iteration an agent $i$ and one of its neighbors $j$ are chosen randomly. Agent $i$ observes agent $j$ state $s_j$ and updates its field $\nu_i$ according to Equation \ref{eq:dynamics}.  Time $t$ is measured in terms of  the number of iterations $n$ averaged over the agents. 

To  consider that $\alpha=\beta$ (and, therefore, $a=b$) is equivalent to the assumption that both the observation $OA$ and the observation $OB$  carry the same weight to the inference process. An observation is in order at this point: since $s_i$ depends only on the sign of  $\nu_i$, this field can be rescaled as  $\nu_i^*=\nu_i/a$ with the dynamics depending only on the ratio $b/a$.

In the process of diffusion of an innovation  the likelihoods are clearly  not necessarily equal. The newer an innovation is, the lesser the observation of a non-adopter should weight as evidence against a product \cite{wejnert02}. However, the number of
adopters of a well valued new product is expected to increase as the
innovation becomes progressively better known by the market. Namely,
the weight as evidence against an innovation conveyed by the observation
of non-adopters ($s_{i}=-1$) should increase as a product ages.

Suppose, as before, that $\alpha$  denotes the probability that a neighbor is an adopter given that it is true that adopting is the best choice that can be made ($A$) and  that $\beta$ denotes the probability that a neighbor is a non-adopter given that non-adopting is the best choice ($B$). These probabilities should, however, be conditional on  the consideration of whether the neighboring agent actually tested the innovation or not. 

As adopters have necessarily tested the new product, $a$ can be kept constant. Considering $b$,  it is possible that a non-adopter has never tried the new product. Hence, let us suppose  $\rho$ to be 
the probability that a neighbor has tested the innovation. We can rewrite the likelihood as  $P(OA|A)= P(OA|A,G)\rho + P(OA|A,\lnot G)(1-\rho) $, where $G$=``\textit{the observed agent tested the innovation}''. As the observation of an adopter that has not tested the product is a contradiction we get  $P(OA|A)= \alpha\rho$, where we have assumed that $P(OA|A,G)=\alpha$.

The remaining likelihoods can be easily calculated to give:
\begin{equation}
\begin{array}{ccc}
P(OA|A) & = & \rho\alpha\\
P(OA|B) & = & \rho(1-\beta)\\
P(OB|A) & = & \rho(1-\alpha)+(1-\rho)\\
P(OB|B) & = & \rho\beta+(1-\rho).\end{array}\label{eq:likelihoods}
\end{equation}

When $s_{j}=+1$ is observed ($OA$) $\rho$ is eliminated by Bayes
theorem and Equation~\ref{eq:logchange} holds for $a$, as
expected. For $b$, Equation~\ref{eq:logchange} is replaced by 
\begin{equation}
b=\log\frac{\rho\beta+(1-\rho)}{\rho(1-\alpha)+(1-\rho)}.
\label{eq:likelihoodnoadopt}
\end{equation}

Notice that when $\rho=1$, Equation~\ref{eq:likelihoodnoadopt}
yields  Equation~\ref{eq:logchange}, as it is expected
in the case where an agent assumes that every other agent has tested the innovation. When an agent assumes that the new product has never been tested by anyone, to say, when  $\rho=0$, we find $b=0$, also as expected.

The general message is that agents need an estimate of how the proportion $\rho$ of agents that tried the new product changes with time. Obviously, $\rho$ should increase monotonically
and a simple assumption would be considering a linear growth
from $\rho=0$ at $t=0$ to $\rho=1$ at a time $t=T$. Time $T$ can be regarded as the agents estimate for the duration of a period in which a product can be considered as a novelty (\textit{novelty period}).  We should  be careful  as $\rho=0$ at $t=0$ would imply, due to the update Equation \ref{eq:bayes}, a certainty that non-adoption is the best choice that can be made. This problem is, however, dealt with by the introduction of log-odds.

Thus we may define a schedule for $\rho$ as
\begin{equation}
\rho(t)=\left\{ \begin{array}{cc}
\frac{t}{T}, & \textnormal{if }t\leq T\\
1, & \textnormal{if }t>T.\end{array}\right.\label{eq:rhot}
\end{equation}

\begin{figure}[htp]
\hspace{-0.5cm}
\includegraphics[width=0.5\textwidth]{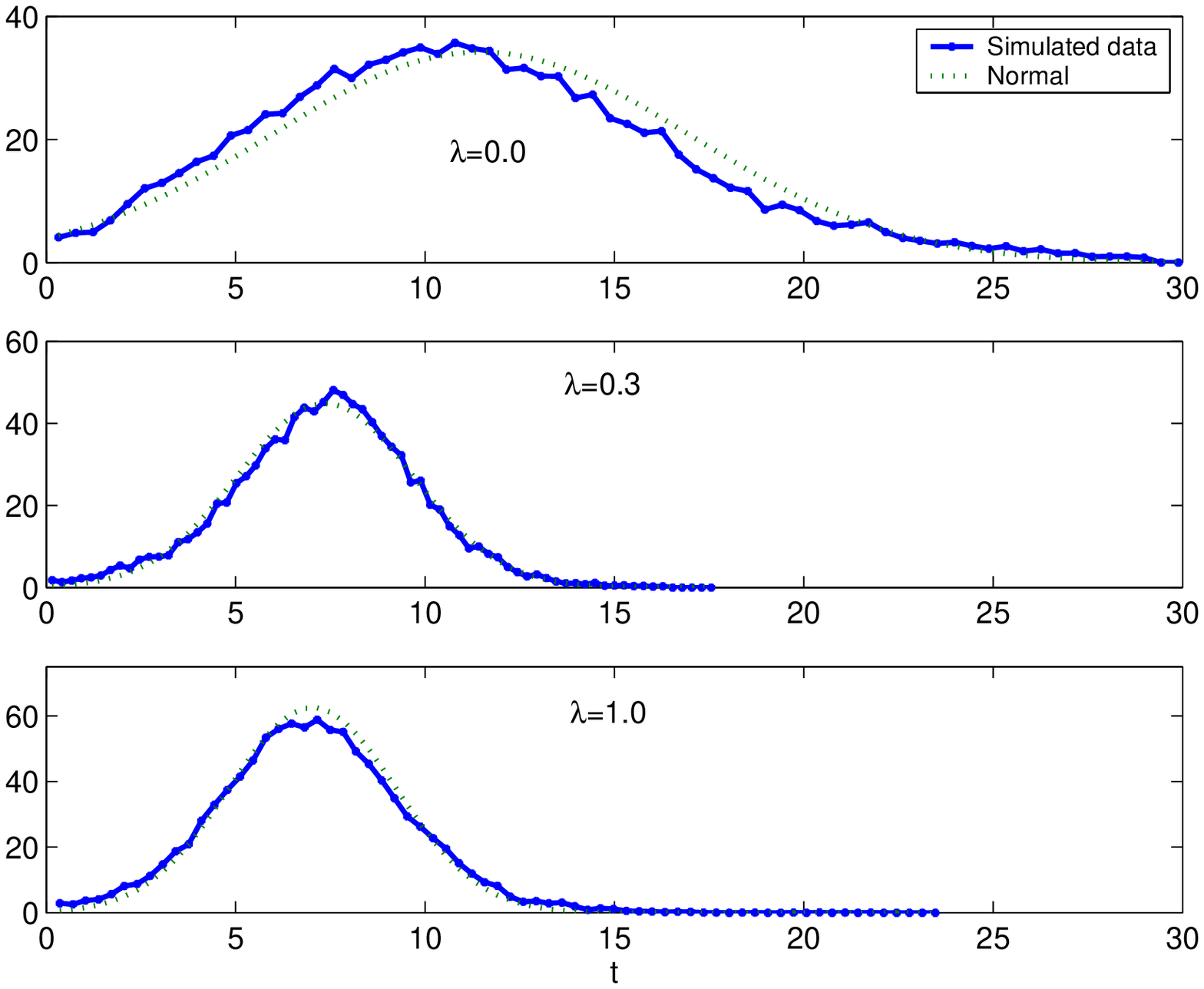}\hspace{0.5cm}\includegraphics[width=0.51\textwidth]{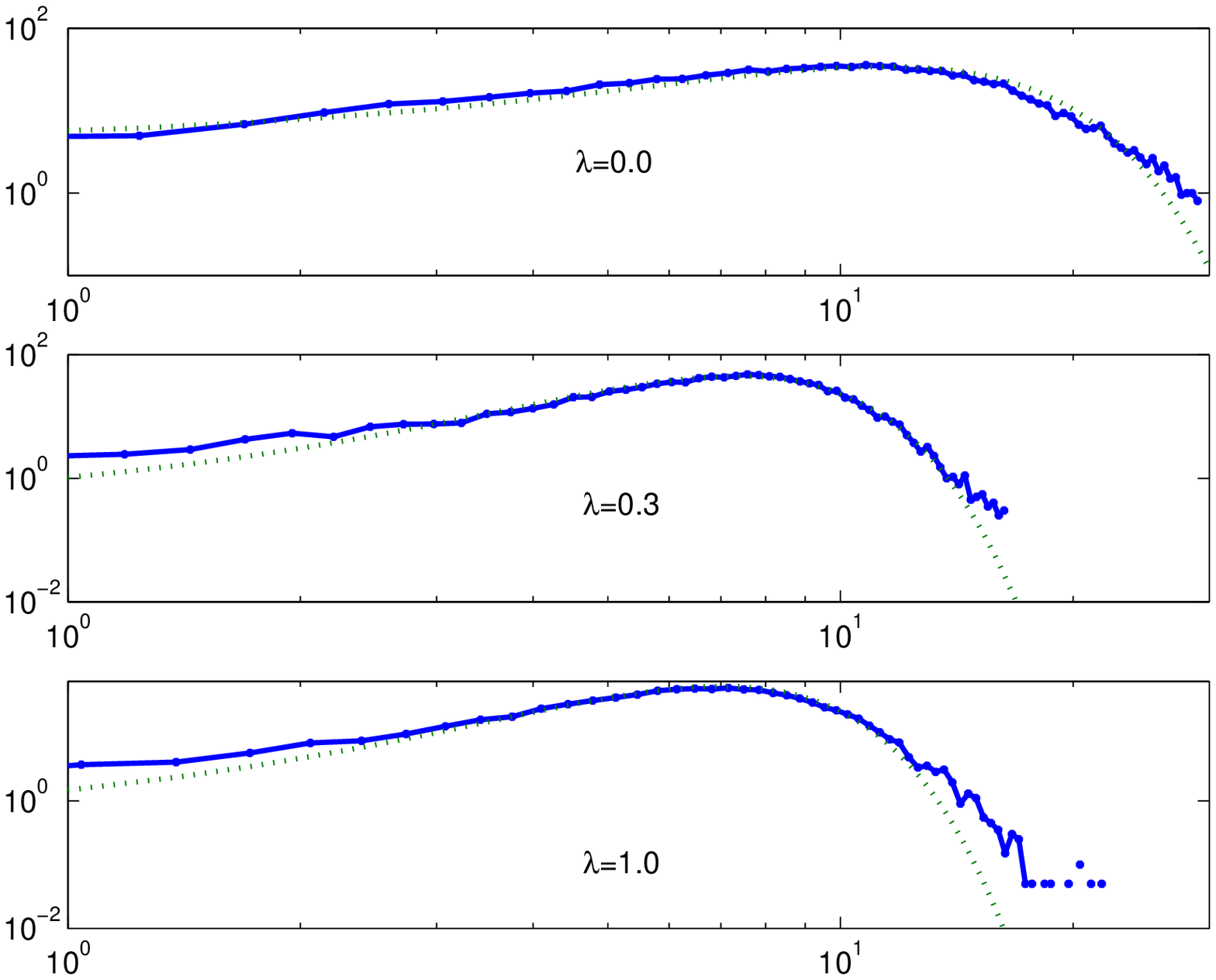}
\caption{ \textit{(Left Panel) Distribution of first adoption times:}  Distribution of first adoption  times as compared to a normal
distribution with the same mean and variance, for a novelty period duration $T=75$ (in lattice steps). Simulations shown correspond
to a single run in a 2D lattice with $32^{2}$ sites (agents), periodic boundary conditions, and rewiring probabilities of $\lambda=0.0$, $0.3$ and $1.0$. Initial adopters are chosen to be $1\%$ of the population and are placed at random sites and $\rho$ is updated at each asynchronous iteration. \textit{(Right Panel) Distribution tails}: The same plots of the left panel in a log-log scale. Long time tails are clearly heavier than they are in a normal distribution.}
\label{fig:spread32} 
\end{figure}

Even if the actual $\rho(t)$ changed in a more complex way, agents could still decide by supposing  such a schedule. The main idea here is to provide the agents with a world model as simple as possible.  Another possibility  would be considering the use of the actual fraction of agents that tested the innovation as an estimate for $\rho(t)$. This alternative would be, however,  rather unrealistic since, in the absence of an external field, agents should only  process local information.

\section{Simulation results}

A series of simulations has been performed to study the variation of
the CODA model proposed above. The dynamics has been run until it reaches  a stable state 
around the novelty period duration $t=T$, when  adopters and non-adopters influence each other with equal weight
and the dynamics becomes that of the CODA model,
with  reinforcement  of opinions within homogeneous domains.
Since the topology of the social network is probably relevant to the dynamics,
simulations have been run with a regular square lattice
and with a random lattice generated by rewiring links with probability $\lambda$.

Initial conditions have been chosen so that a very small
proportion of agents are early adopters (typically, between
0.1\% and 1.0\%, depending on the number of agents). In order to study
the influence of the location of early adopters we have simulated  two scenarios: 
1. setting random sites as adopters ($s_i=+1$), while keeping the remaining sites  as non-adopters (random). 2.  setting a random sequence of neighboring sites as adopters (cluster).

The left panel of Figure~\ref{fig:spread32} shows the distribution of first adoption times  for a single run of size $N=32^{2}$ agents in a regular lattice
($\lambda=0$) and for $\lambda=0.3,1.0$. The simulations depicted have been performed with the novelty period duration set to  $T=75$ (in lattice steps). Initial adopters have been chosen to be $1\%$ of the population and have been placed at random sites. The fraction of testers $\rho$ has been  updated at each iteration following Equation \ref{eq:rhot}. The general shape of the distribution matches, as expected, that of Rogers' normal. A first period with few people adopting the innovation is followed by a period when  adopting is the typical choice and, finally, by a few late adopters being reached. In contrast with
the marketing literature,  no \textit{ad-hoc} classification of consumers (agents) into segments has been employed. Except for the initial random choice of
a few early adopters, every agent behaves in an homogeneous manner.

While,  the normal distribution seems to be a good description of the observed behavior, a closer look 
shows that late adopters are more common than predicted by Rogers' proposal.
That it is so can be clearly verified in the right panel of Figure~\ref{fig:spread32},
where the same curves have been plotted in  log-log scales.  Finding a fat tailed behavior for  long first adoption times actually seems to be more reasonable than a Gaussian decay, as it seems plausible that consumers might still be willing to  adopt an innovation many standard deviations away from the mean adopting time. In Section 4 we provide empirical evidence for this assertion.

\begin{figure}[bbbh]
\includegraphics[width=0.5\textwidth]{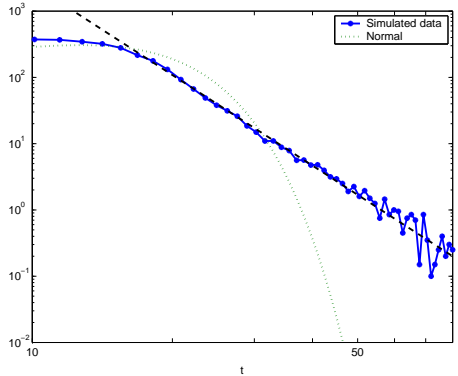}
\caption{\textit{Heavy tailed distribution of first adoption times:} Distribution  of first adoption times as  compared to a normal
distribution with the same mean and variance. Simulations are for a system of size
$64^{2}$ agents. In order to allow a slower adoption, we have taken $T=225$ as the duration of the novelty period. Initial adopters correspond to $1\%$ of the population spread randomly throughout a 2D lattice. A power law has been fit to the tail distribution.}
\label{fig:spread6loglog} 
\end{figure}

In order to examine still closer  large first adoption times,  we have run 
simulations  for a larger system of size $N=64^{2}$ agents. The distribution of first adoption times  for this case 
is shown in Figure~\ref{fig:spread6loglog}, where we have taken $T=225$. A power law has been fit to the tail, in even  clearer  contrast to the usual assumption of normality.

Another noticeable feature of Figure~\ref{fig:spread32} is the dependence of first adoption times on the mean distance between agents in a social network. As  $\lambda$ increases, the social network changes from a regular lattice into a small world network and, finally, into a random graph.  A larger $\lambda$  implying  shorter mean distance and also shorter mean first adoption time.

\begin{figure}[htp]
\hspace{-0.5cm}\includegraphics[width=0.5\textwidth]{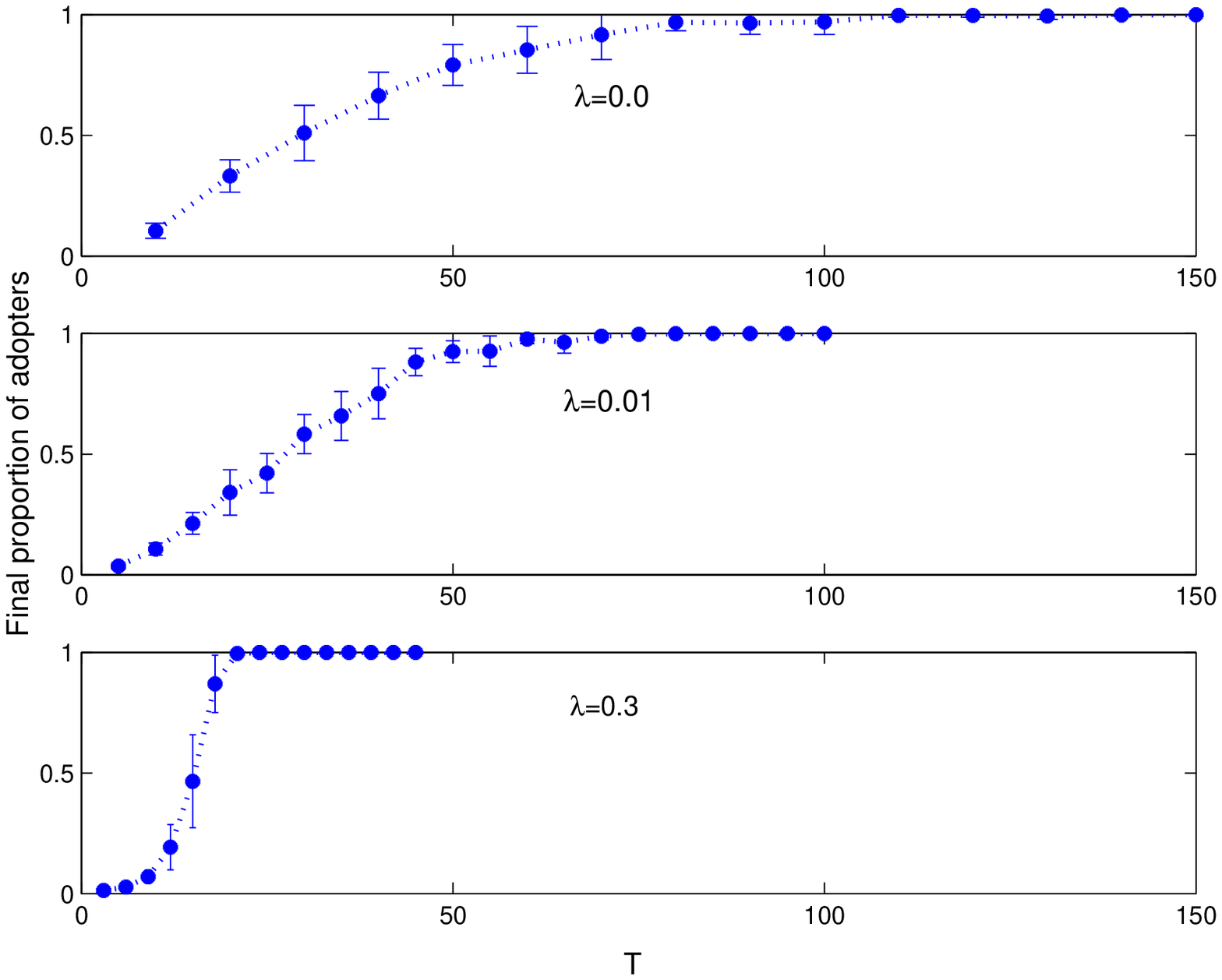}\hspace{0.5cm}
\includegraphics[width=0.51\textwidth]{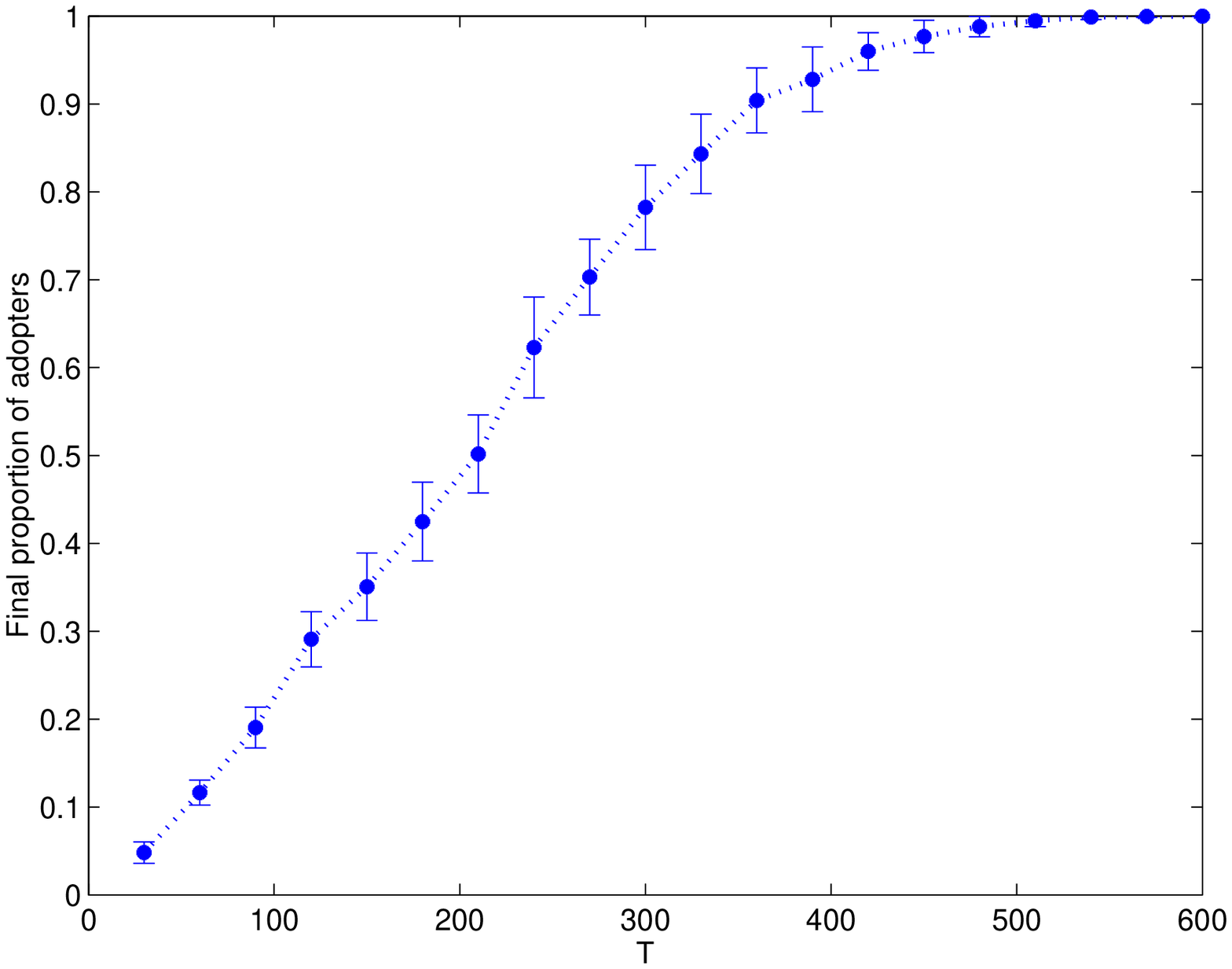}
\caption{\textit{Final proportion of adopters as a function of the novelty period  duration $T$:} The mean and standard deviation of the final proportion of adopters after $20$ runs of a system of size $N=64^{2}$ are depicted. (\textit{Left panel}) Initial conditions are prepared to have $0.01\%$ of the population as randomly spread adopters in lattices with rewiring probabilities of $\lambda=0,0.01$ and $0.3$. (\textit{Right panel})  The same fraction of the population is set to a cluster of  initial adopters in a regular 2D lattice ($\lambda=0$).}
\label{fig:timeadopspreadn64} 
\end{figure}

As opinions are reinforced the diffusion process may freeze before the novelty period duration $T$ is actually reached and a fraction of non-adopters 
may  survive. The panels of Figure~\ref{fig:timeadopspreadn64} depict the  proportion of adopters  for initial adopters spread randomly  throughout the network (left panel) and in an initial cluster (right panel). Results shown
correspond to averages over $20$ runs with one standard deviation wide error bars.
In both figures  durations of the novelty  period  $T$ are in  terms of lattice steps.   The final proportion of adopters grows with $T$.

The left panel of Figure~\ref{fig:timeadopspreadn64} shows that even a small rewiring probability $\lambda=0.01$ leads
to noticeable effects on the final proportion of  adopters. As $\lambda$
grows, the mean distance decreases, diffusion becomes
easier, yielding an increased proportion of adopters. For  $\lambda$ large enough a qualitative  change  is observed with  an abrupt transition  from a state dominated by non-adopters to a
  state dominated by adopters appearing  at a specified $T$.

The right panel of Figure~\ref{fig:timeadopspreadn64} shows the case of
an initial cluster of adopters in a 2D square lattice ($\lambda=0$). The  process of innovation diffusion  from an initial cluster is much slower than in the case of randomly spread adopters. This observation may suggest that in order to reach a larger proportion of adopters it should be favored to launch an innovation at locations as diverse as possible.

\begin{figure}[htp]
\hspace{-0.5cm}\includegraphics[width=0.5\textwidth]{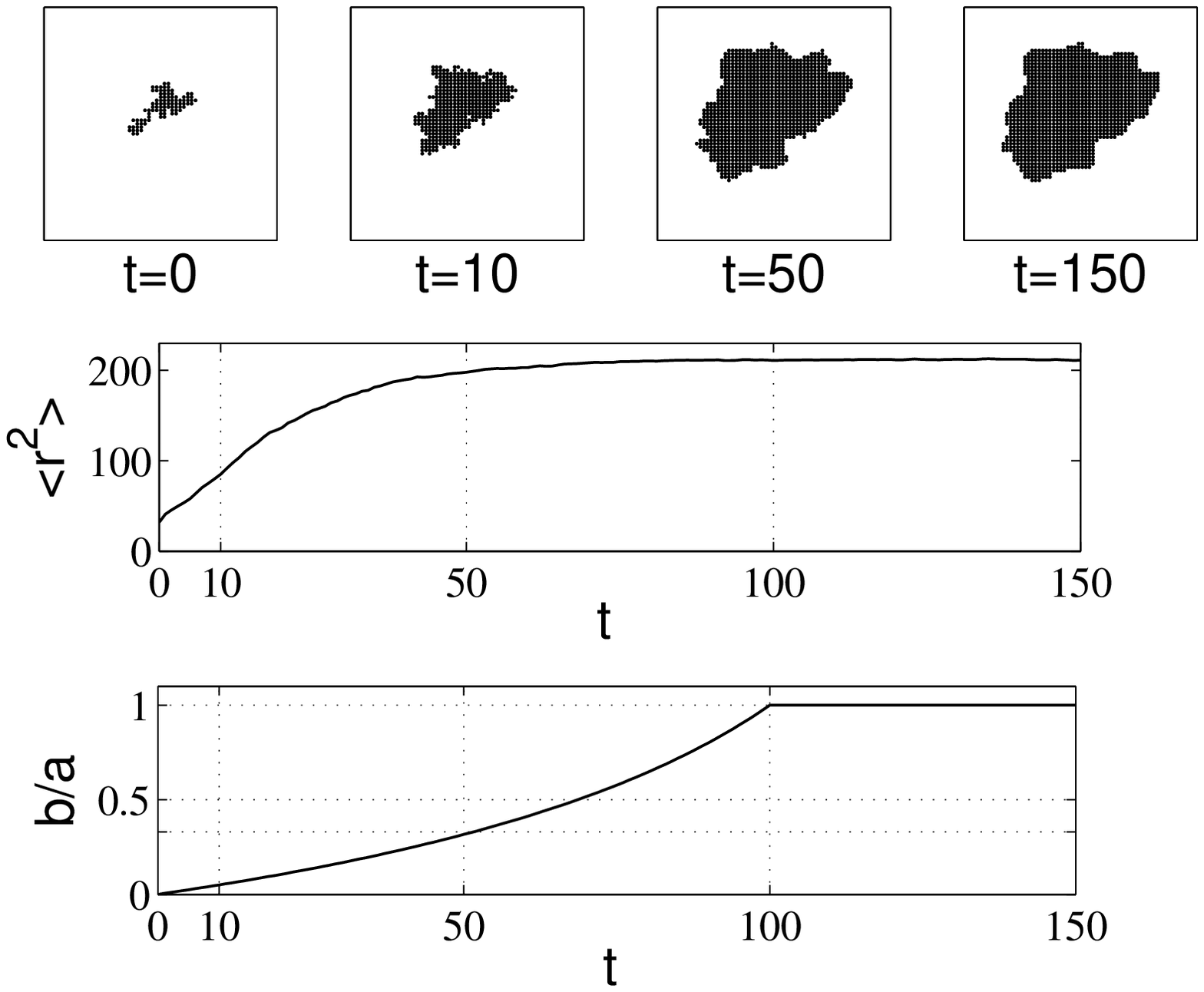}\includegraphics[width=0.6\textwidth]{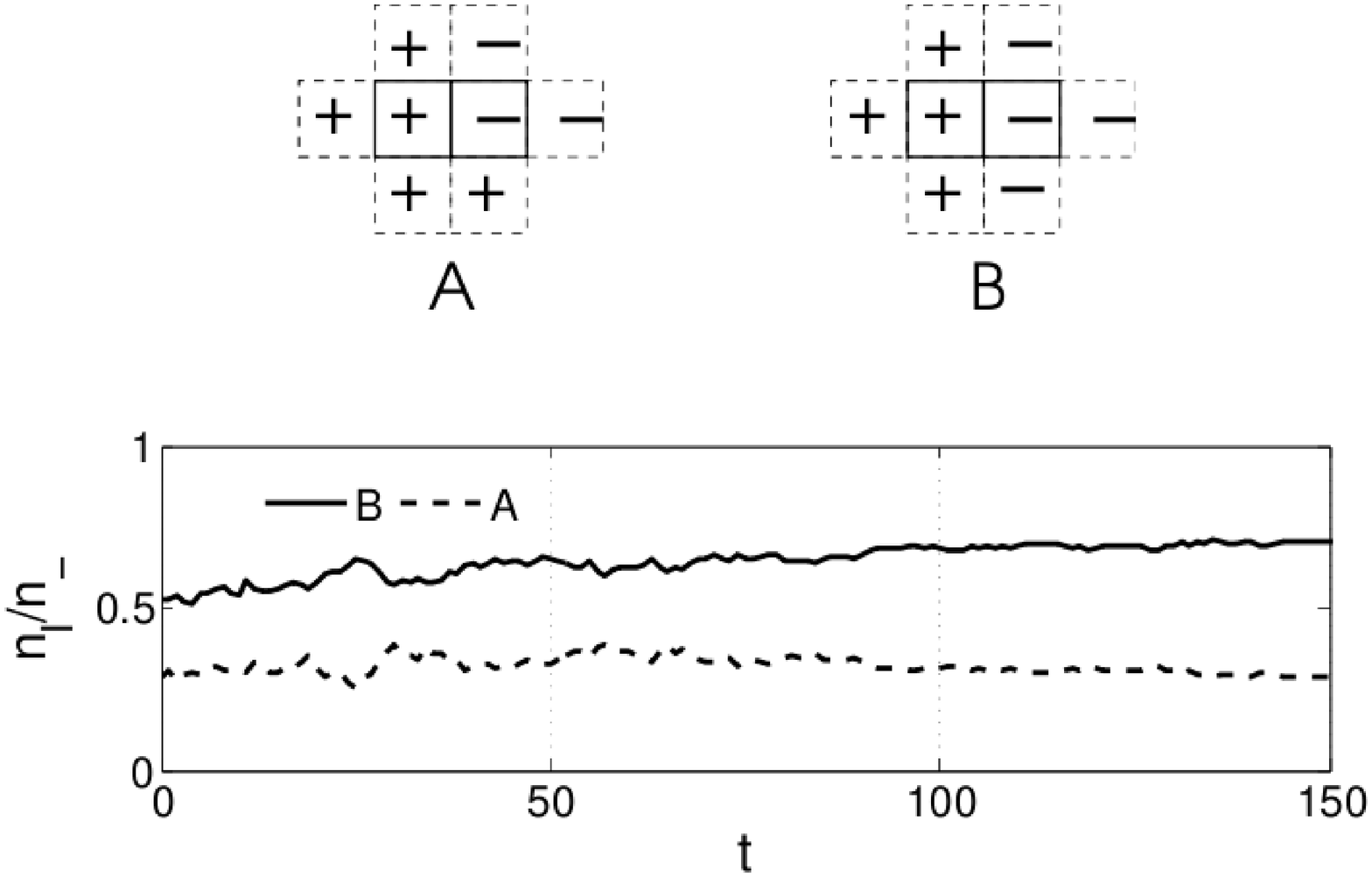}
\caption{\textit{(Left) Diffusion process:} This figure depicts the result of a single run for a system of size $N=512^2$ in a regular 2D lattice and starting from a cluster of adopters containing $1\%$ of the sites. The duration of the novelty period is set to $T=100$. (\textit{Top}) Illustrative view of the growth process. The diffusion is already stalling as frozen interfaces appear at $b=a/3$, corresponding to $t=50$ in the figure. (\textit{Center}) Average  of the squared distance between adopting sites and the center of mass of the adopting domain. The diffusion coefficient decreases as the duration of the novelty period $T=100$ approaches.  (\textit{Bottom}) At each iteration, log-odd fields are modified by $b$ if a non-adopter is observed. 
\textit{(Right) Interfaces:} For  $b<a/3$ ($T<50$) adopting domains grow unhindered. For $b>a/3$ interfaces of type $B$ freeze as both adopter and non-adopter are reinforced. At $b=a$ interfaces of type $A$ are frozen as the adopter is reinforced while the log-odd field  drift for the non-adopter  vanishes. The  bottom right panel shows the fraction $n_I/n_{-}$ of interfaces of each kind, where $n_-$ is  the total number of non-adopting sites in interfaces and the number of interfaces of each type is represented by $n_I$.}
\label{fig:coef_dif} 
\end{figure}

In order to quantify the diffusion process we choose as initial configuration a small cluster of adopters ($s_j=+1$) inside a non-adopting homogeneous phase and study the growth of the adopting phase by coarsening. We proceed by attaching an Euclidian metric to the lattice and computing the  average over adopters of their square radius to the center of mass  $\langle r^2 \rangle(t)$ as it evolves with time. The center left panel of Figure \ref{fig:coef_dif} depicts the growth law yielded. The growth process can be regarded as a diffusion that slows down following a particular prescription for $\rho(t)$. The top left panel of Figure \ref{fig:coef_dif} shows a domain of adopters as it grows with time. The survival of non-adopters can be explained by the  diffusion process becoming progressively slower with time.   The bottom left panel of Figure \ref{fig:coef_dif}
shows the evolution of the modification $b$ in the local fields as a non-adopter is observed. For $b<a/3$ the local field at a site in the center of any heterogeneous neighborhood drifts towards adoption and the domain grows unhindered. The adoption process takes place at interfaces, namely, at non-adopting sites with at least one neighboring adopter. On a 2D regular lattice an interface can be of four types depending on the number of nighboring adopters. We call type $A$ a non-adopting site with two neighboring adopters, type $B$ when a single neighboring adopter is present,  type $C$ when there are three neighboring adopters and type $D$ when all neighbors are adopters.  We define $n_-$ to be the total number of non-adopting sites in interfaces. The number of interfaces of each type is represented by $n_I$, with $I=A,B,C,D$. The right panel of Figure \ref{fig:coef_dif}  shows relative frequencies $n_I/n_-$ of the two main kinds of interfaces observed in a 2D regular lattice. For $b>a/3$,  $B$ type interfaces, that represent more than $50\%$ of all interfaces, freeze as both the adopting and non-adopting sites have their local fields reinforced.  At $b=a$, $A$ type interfaces freeze as the adopting site field is reinforced and the non-adopting field experiences a vanishing drift. For sites surrounded by homogeneous neighborhoods, local fields $\nu_i$ increase by mutual reinforcement,  homogeneous domains are formed and  opinion changes become progressively more difficult with time.

All in all, the  model we have proposed may actually suggest two practical product launching strategies. If the main goal is reaching all  consumers quickly, launching and communicating tasks have to be as wide as possible throughout a social network. If the goal is instead reaching just a
few selected consumers, launching and communicating have to be more localized.
It should be noted, however, that launching may fail completely for
$T$ small  and randomly spread agents. The same is not observed if launching is initiated
from a cluster, since  opinions are mutually  reinforced from the beginning, thus adding the virtue of reduced failure risk to the second strategy.

\section{Comparison to empirical data}

\begin{figure}[!htp]
\includegraphics[width=0.6\textwidth]{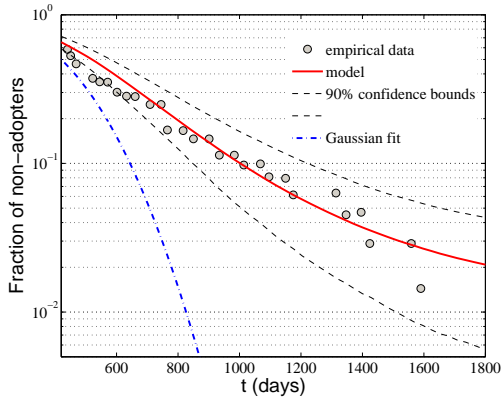}
\caption{\textit{Empirical data:} Grey circles represent data for the fraction of non-adopters of laparoscopic cholecystectomy technologies within a population of $N=59$ hospitals from its inception in 1991 to 1998 (following  \cite{poulsen98}). As we are solely concerned with late adopters only data for times above the the median   $t_{med}=420$ days are shown. The (blue) dot-dashed line represents the best fit to data corresponding to times below the median of a cumulative Gaussian distribution with mean equals to $t_{med}$ (best dispersion equals to $\sigma_t=174.7$). In order to perform a non-rigorous comparison to the model we have chosen the lattice step (or the time needed for an average of one iteration per agent to take place) to represent the period of one month ($30$ days). The full (red) line represents the average over $10$ runs in a lattice of size $N=256^2$, rewiring probability $\lambda=0.04$, novelty period of $T=85$ (in months) and a fraction of $1\%$ of innovators located in a initial cluster. Dashed (black lines) depict  $90\%$ confidence intervals, calculated by choosing maximum and minimum in $10$ runs. The curve represented is independent of the lattice size as the fraction of adopters is normalized by the number of final adopters for comparison purposes. The fat tail behavior observed in the simulations is statistically significant and satisfactorily matches the data.
 }
\label{fig:empirico} 
\end{figure}

Despite the drastic simplifications we have imposed by our prescriptions of social structure, interaction patterns and agent behavior, in this section we compare a general prediction of the model to empirical data. In particular, as a first test we seek to compare statistics for late adopting times, where we have found a sharp contrast between model predictions and what is generally believed in the innovation diffusion literature. A survey study on the introduction of a new non-invasive medical diagnostic  technology (laparoscopic cholecystectomy) in Denmark is described in \cite{poulsen98}. This technology was introduced at two Danish hospitals by January 1991, corresponding to $3.4\%$ of a population with  $N=59$ hospitals.  By 1998 an adoption rate of $98\%$ ($58$ adopters)  was observed in that country. In Figure  \ref{fig:empirico} we show as circles empirical data for the fraction of non-adopters as a function of time (in days). As we are interested solely on late adopters, only times larger than  the median ($t_{med}=420$ days) are depicted. The best fit of a cumulative Gaussian distribution with mean equal to   $t_{med}$ and variance estimated for $t<t_{med}$ ($\sigma_t=174.7$)  is shown in the same figure as a (blue) dot-dashed line. In order to perform a non-rigorous comparison to the model we have chosen the lattice step (or the time needed for an average of one iteration per agent to take place) to represent the period of one month ($30$ days). A good heuristic fit to the data has been found by setting parameters to be: $N=256^2$ (lattice size), $\lambda=0.04$ (probability of rewiring), $T=85$ (novelty period in months) and a fraction of $1\%$ of innovators. In Figure  \ref{fig:empirico} we show a $10$ runs average as a full (red) line. A $90\%$ confidence interval, calculated as the maximum and minimum curves in $10$ runs, is also shown as dashed (black) lines. It is important to observe that the general curve observed is independent of the lattice size as the fraction of adopters used is normalized by the number of final adopters for comparison purposes. The fat tail behavior observed in the simulations is statistically significant and satisfactorily matches the data.

\section{Conclusions}

Before concluding, two further comments are in order. Firstly, agents in the model we have proposed  are not fully rational as  the inference rules employed are just an approximation of the full Bayesian analysis  that should be employed to extract all information contained in the observation of neighboring choices.  Allowing likelihoods to change in time  can actually be regarded as a step towards the perfect foresight required by agents to be fully rational. Secondly, several discrete models can be seen as special cases of the same dynamics~\cite{galam05b}, however, the variation of the CODA model we have analyzed  belongs to a more general class of models based on Bayesian rules and  that includes continuous models~\cite{martins08e}.

We have been able to show that  Rogers' normal curves 
can emerge without the need for presuming {\it ad-hoc} consumer behavioral heterogeneities, such that the  
introduction of  ``opinion leaders''. It is apparent instead that
market segments described in the literature on marketing (innovators, early
adopters, early majority, late majority and laggards) are unnecessary  to explain the emergence of such ``normal curves''. While heterogeneous agents can be behind the observed adoption curve,
it is  clear that the assumption of homogeneous agents can also yield a similar behavior.

An interesting feature of the model we have studied is that the diffusion coefficient decreases 
with time, vanishing as the system reaches a stable state at $T$. For $t>T$, the
model reproduces the CODA model with extremist clusters emerging. 
In that case, interfaces between adopters and non-adopters freeze. 
A number of functional shapes for $\rho(t)$ in Equation~\ref{eq:rhot}
can still be tried in further work and might lead to a  distinct dynamics for the adopting domain growth.

Although we have chosen to employ the marketing literature parlance, any new idea spreading throughout a model society would be expected to exhibit a similar behavior as we have described, given that no external agent (e.g. an advertising campaign or the possibility of objective experimentation) is introduced.
For times $t<T$, the model predicts that  new ideas can spread very easily, however, as novelty effects vanish, domains of supporters and non-supporters and extremely confident opinions emerge.

Finally, a fat tailed distribution for late adopting times emerges in this model. Since it makes
sense that  a few agents may become late adopters even several standard deviations away from the mean time, we believe that to represent a more accurate  description of reality. We test this feature of the model against empirical data to find a statistically significant match.  Further investigation on  quantitative  modeling of the innovation diffusion  phenomena is clearly still  necessary. Nevertheless, considering the degree of simplification employed,  we have found the first results here reported to be encouraging.

\section{Acknowledgement}

This work has been funded by Funda\c{c}\~ao de Amparo \`a Pesquisa
do Estado de S\~ao Paulo (FAPESP), under
grant 2008/00383-9 (ACRM) and  by Conselho Nacional de Desenvolvimento Cient\'{\i}fico e Tecnol\'ogico (CNPq), under grant 550981/2007 (RV). We would like to thank Francisco Javier Sebastian
M. Alvarez for a number of truly relevant insights.

\bibliographystyle{elsarticle-num}
\bibliography{diffusion}

\end{document}